

\input harvmac
\input epsf

\def\exampletree{\epsfxsize=.6cm\epsfbox{tree.eps}}
\def\gstr{\gamma_{\rm str}}

\def\myappendix{\global\meqno=1\global\subsecno=0\xdef\secsym{\hbox{}}
\bigbreak\bigskip\noindent\centerline{\bf Appendix}\message{(Appendix)}
\writetoca{Appendix}\par\nobreak\medskip\nobreak}

\lref\me{M. Wexler, ``Matrix models on large graphs,''
May 1993, Princeton preprint PUPT-1398, hep-th/9305041,
to be published in {\it Nucl. Phys. B}.}
\lref\meone{M. Wexler, ``Low temperature expansion of matrix models,''
March 1993, Princeton preprint PUPT-1384, hep-th/9303146.}
\lref\BIPZ{E. Br\'ezin, C. Itzykson, G. Parisi, and J.-B. Zuber,
{\it Commun. Math. Phys.} {\bf 59} (1978) 35.}
\lref\Harary{F. Harary, {\it Graph Theory} (Addison-Wesley, Reading,
Mass., 1969).}
\lref\ADFO{J. Ambj\o rn, B. Durhuus, J. Fr\"{o}hlich and
P. Orland, {\it Nucl. Phys.} {\bf B270} [FS16] (1986) 457.}
\lref\Kazakov{V. Kazakov, {\it Mod. Phys. Lett.} {\bf A4} (1989) 2125.}
\lref\BKKM{D.V. Boulatov, V.A. Kazakov, I.K. Kostov and A.A. Migdal,
{\it Nucl. Phys.} {\bf B275} [FS17] (1986) 641.}
\lref\DurAmb{B. Durhuus, unpublished; J. Ambj\o rn, ``Barriers in
quantum gravity'' (talk presented at the ICTP Spring School and
Workshop on String Theory, Trieste, April 1993), Niels Bohr Institute
preprint NBI-HE-93-31.}
\lref\MAMA{S.R. Das, A. Dhar, A.M. Sengupta and S.R. Wadia,
{\it Mod. Phys. Lett.} {\bf A5} (1990) 1041.}
\lref\Korch{G.P. Korchemsky, ``Matrix model perturbed by higher curvature
terms,'' April 1992, Parma preprint UPRF-92-334.}

\Title{\vbox{\baselineskip12pt\hbox{PUPT-1412}
\hbox{hep-th/9307120}
}}
{\vbox{\centerline{Critical Behavior of Target Space}\vskip5pt
	   \centerline{Mean Field Theory}}}
\centerline{{Mark Wexler}\footnote{$^\dagger$}
{E-mail address: \it{wexler@puhep1.princeton.edu}}}
        \vskip2pt\centerline{\it{Department of Physics}}
        \vskip1pt\centerline{\it{Princeton University}}
        \vskip1pt\centerline{\it{Princeton, NJ 08544 USA}}
\rm
\vskip .5in
\noindent
Recently, the free energy of the target space
mean field (TSMF) matrix model has been calculated
in the low temperature phase, order-by-order in
a low temperature expansion.  The TSMF model is
a matrix model whose discrete target space has
an infinite coordination number, and whose free
energy assumes a universal form, corresponding
to baby universes joined into a tree.  Here the
free energy is summed to all orders, and expressed
through a transcendental algebraic equation, using
which we analyze the critical phenomena that occur
in the TSMF model.  There are two critical curves,
at which the matter and the geometry become
critical, and for which the critical exponents
are $\alpha = {1\over2}$ and $\gamma_{\rm str}
= -{1\over2}$, respectively.  There is a bicritical
point where the curves meet, at which
$\gamma_{\rm str} = +{1\over3}$.

\Date{July 1993}

Matrix models---convenient representations of two-dimensional
random surfaces and string theory---have run into a brick wall,
the central charge $c=1$ barrier.  The models are exactly solvable
when the target graph is a tree (that is, it has no loops) and
the matter has $c \leq 1$, and almost completely mysterious otherwise.

Recently it has been noticed that matrix models on target graphs
that are regular and whose coordination number $\Delta$ is large
are rather simple, at least in their low temperature phase \me.
These models have target graphs with many loops (for example,
$(\Delta+1)$-state Potts models), and often positive infinite central
charge ($\log_q \Delta$ uncoupled $q$-state Potts models, and
$\Delta/2$-dimensional
lattices).  In particular, the leading $1/\Delta$ term in the free
energy contains contributions from all labeled trees, where each
vertex of each tree is a pure-gravity, open baby universe.
Furthermore, this leading term is universal: it is entirely
independent of the
target graph (and therefore of the matter model or embedding
space).  This is rather reminiscent of mean-field theory on
ordinary lattices, where the system becomes independent of the
particular matter model when the (world-sheet) coordination number
gets large.  For this reason we call the present universal matrix model
``{\it target space} mean field theory'' (TSMF).

We begin by briefly summarizing the results of the previous work \me.
The free energy of the TSMF model is a function of the cosmological
constant $g$ and of a parameter $a$, which depends on $\hat{a}
= e^{-\beta}$, where $\beta$ is the inverse matter temperature
of the matrix model.  For a single matter model, for example,
$a = \Delta\hat{a}$; for multiple matter models, $a=e^{\Delta \hat{a}}
- 1$, where now $\Delta$ is a product of the number of models and
the coordination number of each model.  The free energy was expressed
as a low-temperature expansion, a power series in $a$:
\eqn\lte{
F(g,a) = {1\over a} \sum_{n=1}^\infty {a^n\over n!}
\sum_{T \in {\cal T}_n} V(T)}
where ${\cal T}_n$ is the set of all labeled free trees with
$n$ vertices.  (For graph theoretical definitions and examples,
see, for example, \Harary.)  At this stage, the reader might
worry that $a$ is not small if $\Delta$ is large; this is
compensated by the fact that for large $\Delta$, the critical
temperature $\hat{a}_c \approx \Delta^{-1}$, so that $a_c$
remains finite, and \lte~is valid all the way to the singularity \meone.
For each tree $T$, $V(T)$ is a product of a vertex factor
$p_n(g)$, for each $n$-degree vertex in $T$; for example,
$V(\exampletree) = p_1^3 p_3$.  Thus, for the first few orders,
\eqn\exampleF{\eqalign{
F(g,a) & = p_0 + {1\over2} p_1^2 a + {1\over2} p_1^2 p_2 a^2 +
\left({1\over2} p_1^2 p_2^2 + {1\over6} p_1^3 p_3\right) a^3\cr
& \qquad + \left({1\over2} p_1^2 p_2^3 + {1\over2} p_1^3 p_2 p_3
+ {1\over24} p_1^4 p_4\right) a^4 + \cdots\cr}}

The vertex factors
are just moments of a hermitian one-matrix integral:
\eqn\defpn{\eqalign{
p_n(g) & = \left.{\partial^n \Pi \over \partial \lambda^n}
\right|_{\lambda=0} \cr
\Pi(g,\lambda) & = \lim_{N\to\infty} {1\over N^2} \log
{\int D\mu(\phi) \, e^{-\sqrt{N}\lambda\, {\rm Tr} \phi} \over
\int D\mu(\phi)} \cr}}
where $D\mu(\phi) = D\phi \, e^{-{\rm Tr} U(\phi)}$,
$U(\phi)$ is a non-gaussian matrix-model potential, and $\phi$ is
an $N \times N$ hermitian matrix.  In this
work we will treat the cubic potential
\eqn\defU{
U(\phi) = {1\over2}\phi^2 + {g\over\sqrt{N}} \phi^3}
which generates triangulated surfaces, and at the end mention
other, higher-degree, potentials.
The generating function $\Pi(g,\lambda)$ can be evaluated
by the methods of \BIPZ, for instance; we will give explicit
expressions for it below.
We shall use the following notation for the partial derivatives
of $\Pi$:
\eqn\derpinot{
\Pi_1(g,\lambda) = {\partial \Pi(g,\lambda)\over\partial\lambda},\qquad
\Pi_2(g,\lambda) = {\partial^2\Pi(g,\lambda)\over\partial\lambda^2},
\quad {\rm etc.}}

The free energy series \lte~is the leading, $\Delta^0$ term
in the $1/\Delta$ expansion, and is a great simplification
over the general low-temperature series \meone.
The subleading terms (see \me~for examples) are neither
tree-like nor universal.

In \me~we gave closed-form expressions for each order in the
low-temperature expansion \lte.  In order to analyze the
critical behavior of the free energy, what is needed is a
sum of the entire series.  To calculate it, we introduce
the scalar integral
\eqn\defz{
z(p_i,a,\theta) = {\int_{-\infty}^\infty dx \,\,
e^{-\theta\left[{1\over2}x^2 - a(p_0 + p_1 x + {1\over2}p_2 x^2
+ {1\over6} p_3 x^3 + \cdots)\right]}
\over \int_{-\infty}^\infty dx \,\, e^{-{1\over2}\theta x^2}}}
This is a generating function for closed graphs with labeled vertices of all
degrees, a degree $i$ vertex receiving a factor $p_i,$\foot{We
need to divide $p_i$ by $i!$, lest each line out of each vertex
be individually labeled.} and the graph with $n$ vertices and $\ell$
lines receiving a factor $a^n \theta^{n-\ell}$.  The highest possible
power of $\theta$ will be $\theta^1$, which will multiply trees.
By taking the logarithm and the $\theta\to\infty$ limit, and then
identifying the coefficients $p_i$ with the vertex factors of \defpn,
we obtain exactly the TSMF free energy:
\eqn\Fintegral{
F(g,a) = {1\over a} \lim_{\theta\to\infty} {1\over\theta} \log z(a,\theta)}
We have thus integrated out the hermitian matrices of
the multi-matrix model, leaving behind a single scalar.
In the original matrix model, each vertex represents a point
on a surface; in the effective theory defined by \defz~and \Fintegral,
a vertex represents a baby universe.

Two comments are in order here.  First, the $F$ in \Fintegral~is not
really a sum of the series \lte, but a function whose low temperature
expansion coincides with \lte.  The free energy is expected to be
analytic at $a=0$.  Therefore we can trust \Fintegral~up until a
singularity.  Thereafter, as we shall see below, the $\theta\to\infty$
limit is poorly defined, as the action no longer has a saddle point.

Second,
it would be exceedingly pleasant if the subleading terms in $1/\theta$
in \Fintegral~corresponded to the subleading terms in $1/\Delta$ in the
free energy of the matrix model.  As we pointed out in \me, the latter
are no longer universal; but perhaps some matrix model could be found
for which the subleading---and not just the leading---terms in the
$\theta$ and $\Delta$ expansions coincide, and for which, therefore,
the free energy in \Fintegral~but {\it without} the limit would be
the exact free energy, to all orders in $1/\Delta$.  Unfortunately,
such a matrix model is not likely to exist.  This is because in the
limit $\Delta\to\infty$, the baby-universe vertex factors $p_n$
\defpn~include contributions from twisted surfaces (see \me~for an
example).  As soon as we add loops to the tree, some of the configurations
with twisted vertices are suppressed, as they result in non-spherical
surfaces; therefore the vertex factors become dependent on the global
properties of the tree, and cannot be expressed by means of a local
theory such as \defz, \Fintegral.  Perhaps this difficulty could be
circumvented by taking the double-scaling limit, in which all genera
are present.

Now, back to the calculation.
We evaluate \Fintegral~by steepest descents, of course.  Our final
expression for the free energy is
\eqn\finalF{
a\,F(g,a) = a\,\Pi(g,y(g,a)) - {1\over2} y(g,a)^2}
where the function $y(g,a)$ must be a local extremum which
{\it maximizes} the free energy:
\eqn\TE{
a \,\Pi_1(g,y(g,a)) - y(g,a) = 0}
In the same way that $F(g,a)$, when expanded in powers of $a$, generates
free trees, $y(g,a)$ generates {\it planted} trees, that is trees with
one chosen, degree-one vertex.  We will call \TE, our ``equation of
motion,'' the {\it tree equation}.  The low temperature expansion of
$y$ begins:
\eqn\expy{
y(g,a) = p_1 a + p_1 p_2 a^2 + \left(p_1 p_2^2 + {1\over2}p_1^2 p_3
\right) a^3 + \cdots}

Equipped with the convenient form \finalF~of the free energy, let us
examine its ``physically relevant'' critical regime(s).
Let us first recall the general features of the critical behavior of the
$c<1$ models.  There, the surface does not yet break up into baby
universes, and the temperature is coupled to the order parameter of
the matter.  Given a generic value of the temperature, there is a
critical value of the cosmological constant which makes the area
of the surface diverge; generically, $\gstr = -{1\over2}$.  There is,
in addition, one bicritical point, at which the matter also
undergoes an ordering phase transition, which modifies the geometric
transition, so that $\gstr$ is between $-{1\over2}$ and 0.

In the case of the TSMF model, we expect somewhat different behavior \me.
The cosmological constant $g$ is still coupled to the area of the
surface, but as it only appears in the vertex factors, it is coupled
to the area of the {\it baby universes}, rather than the entire surface.
The (transformed) matter temperature $a$ is no longer coupled to the
``magnetization'' of the matter; instead, as can be seen from \lte,
it is coupled to the size (number of nodes) of the tree.  In this
sense it becomes a second-quantized cosmological constant, coupled
to the number of baby universes.  We still expect a phase transition,
no longer a matter-ordering, but a tree-growing phase transition, the
point where the number of baby universes diverges.  Thus, there will
be two critical curves in the $(g,a)$ plane: the ``T-curve,'' where
the number of nodes (baby universes) in the tree diverges; and the
``G-curve,'' the geometrical phase transitions, where the area of each
baby universe diverges.  If the curves cross there will be a bicritical
point, at which both the number and the area of the baby universes
diverge.

We define the exponents as closely as possible to the usual conventions.
Let $\delta$ be the perpendicular distance in the $(g,a)$ plane between
some point and the T-curve; then as $\delta \to 0$,
\eqn\defalpha{
F(g,a) \approx {\rm analytic} + \delta^{2-\alpha} + \cdots}
We have chosen the symbol $\alpha$ because the T-curve is a remnant
of the matter-ordering transition.  Similarly for the G-curve: let
$\delta$ be the deviation from it, and define
\eqn\defgstr{
F(g,a) \approx {\rm analytic} + \delta^{2-\gstr} + \cdots}
Of course, at the bicritical point the two exponents coincide.

The first thing to do is to calculate the generating function $\Pi(g,\lambda)$,
which is easily done by shifting the matrix $\phi \to \phi + x 1$,
in order to eliminate the linear term; this gives
\eqn\pithree{\eqalign{
\Pi(g,\lambda) & = -\lambda(x+\lambda) - {1\over2}(x^2-\lambda^2) - g x^3
- {1\over4} \log(1 - 12g \lambda) \cr
& \qquad - \Pi(\Gamma(g,\lambda))\cr
x & = {-1 + \sqrt{1 - 12 g\lambda}\over 6g} \cr
\Gamma(g,\lambda) & = (1 - 12g\lambda)^{-3/4} g\cr}}
and $\Pi(g) = -\Pi(g,0)$.\foot{Note the sign convention: all coefficients
in the expansion of $\Pi(g,\lambda)$ about $(0,0)$ are positive.  Therefore
our free energy at $a=0$ will have a negative sign with respect to the
function $E^{(0)}(g)$ in \BIPZ.}
An explicit expression for $\Pi(g)$ can be found in \BIPZ, and is
reproduced in the Appendix.  The geometric singularity in $\Pi(g)$ occurs at
$g = g_0 = 1/\sqrt{108\sqrt{3}} \sim 0.0732$.

The G-curve is the easiest to calculate.  The transition occurs when
$\Gamma(g,y_0) = g_0$; in other words,
\eqn\ynaught{
y_0(g) = {1\over12g} \left( 1 - \left({g\over g_0}
\right)^{4/3} \right)}
We then solve the
tree equation \TE~to obtain $a_0(g)$.  Let us analyze the critical
region by expanding the tree equation; putting $\Delta a = a_0 - a$,
$\Delta y = y_0 - y$, we get
\eqn\expTE{
u(g) + v(g) \Delta y + w(g) \Delta y^{3/2} + \cdots =
{y_0\over a_0} - {\Delta y\over a_0} + {y_0\over a_0^2}\Delta a + \cdots}
where the functions $u$, $v$, and $w$ can be determined from \pithree~after
a little algebra: see Appendix.  We solve for the critical points
by canceling the analytic parts of \expTE.  Canceling the constant term
\eqn\afirst{a_0 = y_0/u}
gives an explicit form for $a_0$ ($\gamma = g/g_0$):
\eqn\anaught{
a_0(g) = {1 - \gamma^{4/3} \over
1 - 2 \gamma^{2/3} + \gamma^{4/3} + 36 g_0^2\gamma^{2/3}
+15552 \cdot 3^{1/4}\,(3+\sqrt{3})^{-3} g_0^3 \gamma^{2/3}}}
If we also cancel the linear term, that is if
\eqn\anext{a_0(g) v(g) = -1}
then $\Delta y$ will scale as $\Delta a^{2/3}$; otherwise, $\Delta y \approx
\Delta a$, and $y(g,a)$ will be analytic on the G-curve.  Taken with
the condition \anaught, the equation \anext~has a unique solution:
\eqn\bicrit{
g_* = {1\over\sqrt{288}} \sim 0.0589, \quad a_* = 1}
This will be seen to be the bicritical point anticipated above.
At all the other points $(g,a_0(g))$ on the G-curve \anaught, $y$ will
be analytic, while $F$ will be singular; we shall return to this point
below.

Now, the T-curve.  Consider a point $(g,a)$ at which $a\,\Pi_1(g,y)-y$ is
analytic (the singular points have been treated in the preceding paragraph).
By the implicit function theorem we know that $y(g,a)$, as a root of the
tree equation, will be analytic, unless the partial derivative of the
equation with respect to $y$ vanishes.  The following equation, together
with \TE, defines the T-curve:
\eqn\tcurve{
a\,\Pi_2(g,y(g,a)) - 1 = 0}
We solve the transcendental system \TE~and \tcurve~numerically.
A convenient way to do so is to pick some value of $g$, solve
$\Pi_1(g,y)/\Pi_2(g,y) - y = 0$, and plug the value(s) of $y$ into
\TE~or \tcurve~to find $a$.  In addition, \TE~and \tcurve~can be
developed in powers of $g$ near $g=0$; there are two branches of the
G-curve, $a_\pm(g)$.  After a little algebra, we find
\eqn\serapm{\eqalign{
y_\pm(g) & = \pm 1 - 6g \mp {63\over2} g^2 +{\cal O}(g^3) \cr
a_\pm(g) & = \pm {1\over6g} - {3\over2} \mp {9\over4}g + {\cal O}(g^2) \cr}}
Having canceled the first derivative of the tree equation to find
the critical T-curve, we ought to search for a multicritical point
on that curve that would be the result of canceling also the second
derivative.  However, we do not find this type of multicritical point
realized in our system.

This is the phase diagram of the TSMF model:
\vskip 1cm
\centerline{\epsfbox{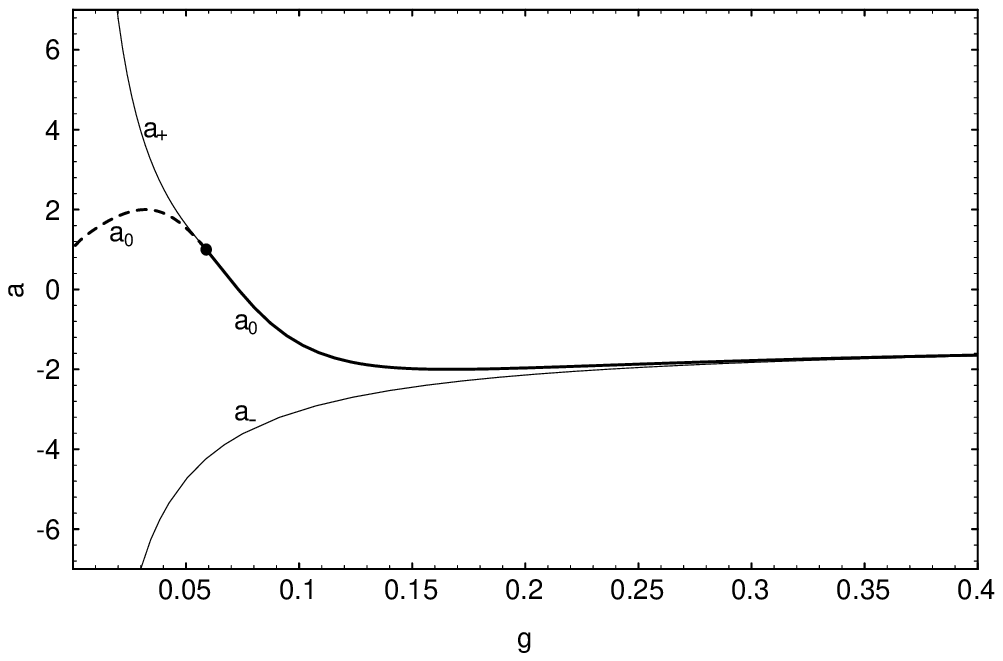}}
\vskip 1cm
As our system is symmetric under $g\to-g$, we only show the right
half.  For completeness we show the $a<0$ quadrant, although only
$a \ge 0$ corresponds to real temperature.
The bicritical point $(g_*,a_*)$ is marked.  It is not only the
solution of \anext, but is also where the upper branch of the T-curve,
$a_+(g)$ meets the G-curve, $a_0(g)$.  For $g>g_*$, there may well
be solutions $a_+(g)$ above $a_0(g)$, but we cannot find them, as
our generating function $\Pi(g,\lambda)$ breaks down in that regime:
matrix models do not make sense above the geometric transition.
$a_-$, on the other hand,
continues indefinitely, approaching $-1$ as $g\to\infty$.  So does
the G-curve, although they never meet; in fact, their difference
$a_0(g) - a_-(g)$ is asymptotically $1/9g$.

Let us consider in more detail what happens to the solutions of the
tree equation \TE~in the different regions of the phase diagram.
First, fix some $g>g_*$.  For $a<a_-(g)$ or $a>a_0(g)$ the
equation has no solutions: our solution for the free energy is
not valid outside this region.  For $a_- < a < a_0$, $y(g,a)$ is
analytic.  Near $a_-$, the number of baby universes on the tree diverges;
$y - y_- \approx \sqrt{a-a_-}$, because we
have canceled one derivative in $y$ of the tree equation.
Of the non-integral powers in $F(g,a)$, the lowest, $\sqrt{a-a_-}$,
is canceled by the tree equation \TE.
Therefore $F(g,a) \approx {\rm analytic} + (a-a_-)^{3/2} + \cdots$, and
$\alpha = {1\over2}$.  The trees diverge as if the baby universes were
not there, as ${1\over2}$ is the generic exponent for tree divergence, found,
for instance, in the ``Cayley'' model (which just counts labeled trees
with nothing at the nodes, in other words with $p_n(g) = 1$),
or in one phase of the toy model of Ambj\o rn, Durhuus,
Fr\"{o}hlich and Orland \ADFO~(in which the role of $a$ is played by
the cosmological constant, and therefore $\gstr = {1\over2}$).
Near $a_0$, the area of each baby universe
diverges.  Since \anext~is not satisfied, $y$ is analytic at
$a_0$.\foot{Strictly speaking it cannot be, since it is not even
defined for $a>a_0$; what is meant is that the series expansion of $y$
for $a<a_0$ in powers of $a_0-a$ includes only non-negative integer
powers.}  However, since $\Gamma(g,y_0) = g_0$, this {\it is} a critical
point for $F$, which, according to \pithree, scales as $(a_0-a)^{5/2}$
+ analytic terms.  Therefore $\gstr = -{1\over2}$: the entire tree of
baby universes diverges as if it were one, empty surface.

Now fix some $g<g_*$.  For $a>a_+$ or $a<a_-$, the tree equation again
has no solutions.  At $a_\pm$ the tree diverges, and $\alpha = {1\over2}$
as before.  When $a$ is between $a_+$ and $a_0$, there are two solutions
to the tree equation; we must take the one which maximizes the free
energy.  By definition, when $a=a_0$, one of these roots is $y_0$,
which satisfies $\Gamma(g,y_0) = g_0$ and so would make the free energy
diverge.  However, this root is a minimum of the free energy, a
``metastable'' state not accessible to the system.  Therefore, when
$g < g_*$, the area of the baby universes never diverges.  (This is
why this part of the G-curve is drawn as a broken line.)

Finally, consider $g=g_*$.  As usual, $\alpha = {1\over2}$ at $a=a_-$.
At $a = a_0 = a_+$, the number of baby universes and their area
diverge simultaneously.  $y$ is no longer analytic; as shown above,
$\Delta y$ scales as $\Delta a^{2/3}$.  The lowest non-integral powers
in the free energy, $\Delta a^{2/3}$ and $\Delta a^{4/3}$, are canceled
by the critical conditions \afirst~and \anext, respectively.  Therefore
$F(g,a) \approx {\rm analytic} + \Delta a^{5/3}$, and $\gstr = +{1\over3}$.

Once again, we stress that there is no magnetization in the TSMF
model.  The T-curve is a tree-growing, not a spin-disordering,
phase transition.  The only exception is the $a=0$ line.  Here,
the $a\to 0$ and $\Delta\to\infty$ limits do not commute.  If
we take the $a\to 0$ limit first, the system will be magnetized
for any value of $\Delta$; taking the limits in reverse order,
in other words putting $a=0$ in the TSMF model, yields no
magnetization.

One could repeat this calculation with a different matrix model
potential $U(\phi)$ than \defU.  As long as the potential is still
at the $m=2$ point \Kazakov, the critical behavior of the model
should be qualitatively the same as that of the cubic model, with
the same exponents, including $\gstr = {1\over3}$ at the bicritical
point.  However, if the potential is tuned to one of Kazakov's
$m \geq 3$ multicritical points, on the T-curve the generic $\alpha$
will still be ${1\over2}$, but on the G-curve the generic $\gstr$
will now be $-{1\over m}$.  At the bicritical point,
$\Delta y \approx \Delta a^{m\over m+1}$, and $\gstr = {1\over m+1}$.

To conclude, we compare our results to related work.  A number of
years ago, several groups studied random surfaces embedded in continuous
$d$-dimensional euclidean space (see \ADFO~and \BKKM, and references
therein).  These models can be viewed as multiple gaussian systems
coupled to a matrix model.  It was argued that in the limit $d\to\infty$,
the surfaces are dominated by branched polymers, that is long thin tubes
connected into trees.  This situation can be viewed as a special case
of the TSMF theory, its regime of small baby universes.  This is not
surprising, as the multiple gaussian models have only one coupling
constant---the cosmological constant---which is coupled to the total
size of the polymer, a role played by the matter temperature in the
TSMF model.  The reader will not be surprised to learn, therefore,
that in certain regimes of the the multiple gaussian systems, the
authors of \ADFO and \BKKM find $\gstr = {1\over2}$, a result analogous,
then, to our $\alpha = {1\over2}$.

Recently, Durhuus and Ambj\o rn \DurAmb~obsereved that if a random surface
model has $\gstr = -{1\over m}$ in its underlying, unbranched
form, its ``polymerized'' version will have $\gstr = {1\over m+1}$.
This is in perfect agreement with our results, with $m$ the multicritical
index.  This is because the trees of empty baby universes, which we
found in \me~to dominate the $\Delta\to\infty$ limit, correspond
precisely to the polymerized surfaces that Durhuus and Ambj\o rn
treat.  There is another model, the $({\rm Tr}\,\phi^2)^2$ model
introduced in \MAMA, which has phases with just the above exponents,
and which Ambj\o rn points out as an explicit realization of the
construction in \DurAmb.  This is because the nonlocal term induces
surface touching, giving surfaces that are trees of baby
universes.\foot{This means that in some sense, in the limit $\Delta
\to\infty$ of multi-matrix models, one can replace the interaction
term in the action by $({\rm Tr}\,\phi^2)^2$.}
The coefficient of the nonlocal term is therefore the analogue of
the matter temperature in the TSMF model, and the phase diagrams
of the two models are quite close, our bicritical point corresponding
to Korchemsky's intermediate phase \Korch.

\bigbreak\bigskip\bigskip\centerline{{\bf Acknowledgements}}\nobreak
The author thanks Alexander Migdal and Marc Potters for encouragement
and useful suggestions, and Jan Ambj\o rn for sending him ref.
\DurAmb~prior to its publication.

\myappendix

Here we give the expansion coefficients for the generating function
$\Pi(g,\lambda)$, required in equation \expTE~and the subsequent analysis.
{}From \BIPZ~we learn that
\eqn\apppi{\eqalign{
\Pi(g) & = -{\sigma\over3}{3\sigma^2+6\sigma+2\over(1+\sigma)(1+2\sigma)^2}
+ {1\over2} \log(1+2\sigma)\cr
\sigma(g) & = -{1\over2} + {1\over 2 \tau(g)} +
{\tau(g)\over6} \cr
\tau(g) & = \left( 3 \left( \sqrt{3(1 - 34992g^4)}
- 324 g^2\right)\right)^{1/3} \cr}}
Near the critical point $g_0 = 1/\sqrt{108 \sqrt{3}}$, $\Pi(g)$ has the
expansion ($\Delta g = g_0 - g$)
\eqn\apppiexp{\eqalign{
\Pi(g) & = \pi_0 + \pi_1 \Delta g + \pi_2 \Delta g^2
+ \pi_{5/2} \Delta g^{5/2} + {\cal O}(\Delta g^3) \cr
\pi_0 & = \sqrt{3} - {3\over2} - {1\over4}\log 3 \cr
\pi_1 & = {144 \cdot 3^{1/4}\over (3+\sqrt{3})^3} \cr
\pi_2 & = -{216 (187 + 108\sqrt{3}) \over 459 + 265 \sqrt{3}} \cr
\pi_{5/2} & = {1152 \sqrt{2} \cdot 3^{3/8} (627 + 362\sqrt{3}) \over
5(362 + 209 \sqrt{3})} \cr}}
We are interested in the expansion of $\Pi(\Gamma(g,y))$ about $y =
y_0(g)$; letting $\Delta y = y_0(g) - y$ and $\gamma = g/g_0$, we get
\eqn\apppiexptwo{\eqalign{
\Pi(\Gamma(g,y)) & = \tilde{\pi}_0 + \tilde{\pi}_1 \Delta y +
\tilde{\pi}_2 \Delta y^2 + \tilde{\pi}_{5/2} \Delta y^{5/2} +
{\cal O}(\Delta y^3)\cr
\tilde{\pi}_0 & = \pi_0 \cr
\tilde{\pi}_1 & = 9 g_0^2 \gamma^{-1/3} \pi_1 \cr
\tilde{\pi}_2 & = -{189\over2} g_0^3 \gamma^{-2/3} \pi_1 +
81 g_0^4 \gamma^{-2/3} \pi_2 \cr
\tilde{\pi}_{5/2} & = 243 g_0^5 \gamma^{-5/6} \pi_{5/2} \cr}}
The rest of the terms in $\Pi(g,\lambda)$ are of course analytic at
$y=y_0(g)$, and their expansion is $\tilde{\phi}_0 + \tilde{\phi}_1
\Delta y + \tilde{\phi}_2 \Delta y^2 + {\cal O}(\Delta y^3)$, where
the coefficients are:
\eqn\appcoeff{\eqalign{
\tilde{\phi}_0 & = {1\over g_0^2} \left( {1\over864 \gamma^2} -
{1\over144\gamma^{2/3}} + {1\over108} - {\gamma^{2/3}\over288}\right) -
{1\over3}\log \gamma \cr
\tilde{\phi}_1 & = -{1\over g_0} \left( {1\over12\gamma} -
{1\over6\gamma^{1/3}} + {\gamma^{1/3}\over12} \right) -
{3g_0\over\gamma^{1/3}} \cr
\tilde{\phi}_2 & = {1\over2\gamma^{2/3}} + {18g_0^2\over\gamma^{2/3}}
-{1\over2} \cr}}
The functions $u(g)$, $v(g)$, and $w(g)$ can now be read off:
\eqn\appuvw{\eqalign{
u(g) & = \tilde{\phi}_1(g) - \tilde{\pi}_1(g) \cr
v(g) & = 2(\tilde{\phi}_2(g) - \tilde{\pi}_2(g)) \cr
w(g) & = -{5\over2} \tilde{\pi}_{5/2}(g) \cr}}

\listrefs
\bye